\documentstyle[12pt,hpatex]{article}

\begin{document}

\title{What Particles Are Described by
the Weinberg \linebreak Theory?\\}

\authors{Valeri V. Dvoeglazov\footnote{On leave of absence from
{\it Dept. Theor. \& Nucl. Phys., Saratov State University,
Astrakhanskaya ul., 83, Saratov\, RUSSIA.}\,
Internet address: dvoeglazov@main1.jinr.dubna.su}}

\address{Escuela de F\'{\i}sica, Universidad Aut\'onoma de Zacatecas \\
Antonio Doval\'{\i} Jaime\, s/n, Zacatecas 98068,
Zac., M\'exico\\ Internet address:  VALERI@CANTERA.REDUAZ.MX
}

\abstract{The main  goal of the paper is to study the origins
of  a contradiction between the Weinberg theorem $B-A=\lambda$ and the
`longitudinality' of  an antisymmetric tensor field (and of  a Weinberg
field which is  equivalent to it), transformed on the $(1,0)\oplus (0,1)$
Lorentz group representation.  On the basis of analysis of dynamical
invariants in the Fock space situation has been partly clarified.}

The interest in the $2(2j+1)$ component
model~\cite{Ahl,Ahl2,Dvoegl,Dvoegl2} and in the antisymmetric tensor
fields~\cite{Avdeev,Avd2} has grown in the last years.  Antisymmetric
tensor fields are of importance for physical applications~\cite{Chizhov}.
Moreover, they are an object of continuous and renewed interests due to
their connection with topological field theories~\cite{Birmingham}.

However, many points are unclear to understand at the moment.
The most  intrigued  thing, in my opinion,
is the following contradiction~\cite{Dvoegl,Dvoegl2}:
the $j=1$ antisymmetric tensor field is shown to possess the longitudinal
components only~\cite{Avd2,Hayashi,Kalb}, the helicity  is equal to
$\lambda=0$.  In the meantime, they transform according to  the  $(1,0) +(0,1)$
representation of the Lorentz group (like a Weinberg bispinor\footnote{See
for the mapping between  antisymmetric tensor field equations and Weinberg
equations  ref.~[3e,4].}).  How is the Weinberg theorem, $B-A=\lambda$,
ref.~\cite{Weinberg},  for the $(A,B)$ representation to be treated in
this case? Moreover, does this fact signify that we must abandon
the Correspondence Principle (in the classical physics we have become
accustomed that antisymmetric tensor field has only transversal
components) ?

In the beginning let me reproduce the previous results.
In ref.~[3e] I worked with the Lagrangian density
(see  also~\cite{Greiner,Santos}):
\begin{equation}\label{eq:Lagra}
{\cal L}^{W}=-\partial_{\mu}\overline\psi\gamma_{\mu\nu}
\partial_\nu\psi-m^2\overline\psi\psi \quad,
\end{equation}
$\gamma_{\mu\nu}$ are the Barut-Muzinich-Williams matrices which are chosen
to be Hermitian.
In  a massless limit,  implying  an  interpretation of the
6-spinor  as\footnote{One can also choose
$$\psi^{(2)}= \pmatrix{\vec E+i\vec B\cr -\vec E+i\vec B} =\gamma_5
\psi\quad.$$
Since $\overline \psi^{(2)} =-\overline \psi \gamma_5$ the formula
(\ref{eq:Lagran}) is not changed.}
\begin{eqnarray}\label{eq:EH}
\cases{
\chi=\vec E+i\vec B\quad,& $ $\cr
\phi=\vec E-i\vec B\quad,& $ $}
\end{eqnarray}
$\psi = \mbox{column} (\chi  \quad \phi)$,  $\vec E$ and $\vec B$ are
the real 3-vectors,
the  Lagrangian (\ref{eq:Lagra}) can be re-written in the  following way:
\begin{equation}\label{eq:Lagran}
{\cal L}^{W}=-(\partial_\mu F_{\nu\alpha})(\partial_\mu
F_{\nu\alpha}) +
2(\partial_\mu F_{\mu\alpha})(\partial_\nu F_{\nu\alpha})
+ 2(\partial_\mu F_{\nu\alpha})(\partial_\nu F_{\mu\alpha})\quad.
\end{equation}
This form of the Lagrangian leads to the Euler-Lagrange equation
\begin{equation}
{\,\lower0.9pt\vbox{\hrule \hbox{\vrule height 0.2 cm
\hskip 0.2 cm \vrule height 0.2cm}\hrule}\,}
F_{\alpha\beta}-2(\partial_{\beta}F_{\alpha\mu,\mu}-
\partial_{\alpha}F_{\beta\mu,\mu})=0 \quad,
\end{equation}
where ${\,\lower0.9pt\vbox{\hrule \hbox{\vrule height 0.2 cm
\hskip 0.2 cm
\vrule height 0.2 cm}\hrule}\,}=\partial_{\nu}\partial_{\nu}$.

The Lagrangian (\ref{eq:Lagran}) is found out  there to be
equivalent to the Lagrangian of  a free massless skew-symmetric field,
given by Hayashi  in ref.~\cite{Hayashi}:\,\footnote{See also for
describing  closed strings
on the ground of this Lagrangian in ref.~\cite{Kalb}.}
\begin{equation}\label{eq:LagHa}
{\cal L}^{H}=\frac{1}{8}F_k F_k \quad,
\end{equation}
with $F_k=i\epsilon_{kjmn}F_{jm,n}$. It is re-written in:
\begin{eqnarray}
{\cal L}^{H}&=&-{1\over 4}(\partial_{\mu} F_{\nu\alpha})
(\partial_{\mu} F_{\nu\alpha})+
{1\over 2} (\partial_{\mu} F_{\nu\alpha})(\partial_{\nu}
F_{\mu\alpha})=\nonumber\\
&=&{1 \over 4}{\cal L}^{W}-{1\over 2}(\partial_{\mu}
F_{\mu\alpha})(\partial_{\nu} F_{\nu\alpha})\quad,
\end{eqnarray}
what proves the statement made above if take into account
the generalized Lorentz condition, ref.~\cite{Hayashi}.
After  applying the Fermi method {\it mutatis mutandis}
as in ref.~\cite{Hayashi}
we  achieved  the  result  that  the Lagrangians
(\ref{eq:Lagra}) and (\ref{eq:LagHa})
describe  massless particles
possessing longitudinal physical components only. Transversal
components are removed
by means of the ``gauge" transformation
\begin{equation}\label{eq:gauge}
F_{\mu\nu}\rightarrow
F_{\mu\nu}+A_{[\mu\nu]}=F_{\mu\nu}+\partial_{\nu}
\Lambda_{\mu}-
\partial_{\mu}\Lambda_{\nu}
\end{equation}
(or by  the  transformation similar to the above but applied to the Weinberg
bivector).
This fact is  very surprising from  a viewpoint  of  the Weinberg
theorem about a connection between the helicity  $\lambda$ and
the Lorentz group representation $(A,B)$ which  field operators
transform on:  \,\,\,  $B-A=\lambda$.

Here I am going to clarify this question. In  ref.~\cite{Dvoegl2}
 the concept of the Weinberg field as a system
of degenerate states has been proposed. Unfortunately,
the consistent description of  the Weinberg ``doubles"~\cite{Dvoegl2}
and/or of the antisymmetric tensor fields $F_{\mu\nu}$ and its dual
$\tilde F_{\mu\nu}$
as the parts of a  degenerate doublet are absent in
the literature (to my knowledge).  Many works
deal with the dual theories, {\it e.g.}~\cite{Strazhev},
but do not contain  quantization topics.

Firstly, we need to choose the appropriate  Lagrangian.
In the case  of the use of the pseudoeuclidean metric (when
$\gamma_{0i}$ is chosen to be anti-Hermitian)
it is possible to write
the Lagrangian following F. D. Santos and H. Van Dam,
ref.~\cite{Santos} (see also ref.~[3b], where this Lagrangian
has been obtained  independently):
\begin{equation}\label{Santos}
{\cal L} = \partial^\mu \overline \psi \gamma_{\mu\nu} \partial^\nu \psi
-m^2 \overline \psi \psi \quad.
\end{equation}

One can use the  Lagrangian  which is similar
to Eq. (\ref{Santos}):\footnote{In  refs.~[3b,13]
the possibility of appearance of the ``doubles"
has not been considered (neither in any other
paper on the $2(2j+1)$ formalism, to my knowledge).}
\begin{equation}\label{eq:Lagran1}
{\cal L}^{(1)} = -\partial_\mu \overline \psi_1  \gamma_{\mu\nu}
\partial_\nu \psi_1
-\partial_\mu \overline \psi_2 \gamma_{\mu\nu}
\partial_\nu \psi_2 - m^2 \overline \psi_1 \psi_1
+ m^2 \overline \psi_2 \psi_2 .
\end{equation}
in the Euclidean metric.
The only inconvenience to be taken in mind where it is necessary
is that we need to
imply that  $\partial_\mu^\dagger = (\vec \nabla, - \partial /\partial x_4)$,
provided that $\partial_\mu = (\vec \nabla, \partial /\partial  x_4)$,
ref.~\cite{Lurie}.

The Lagrangian (\ref{eq:Lagran1}) leads to the equations:\footnote{Problems
related with the Wick propagator and   the Feynman-Dyson propagator
are considered
in the approaching publication in the framework of the proposed model,
see also for the discussion of these topics  on the page B1324 in ref.~[11a]
and ref.~[2].}
\begin{eqnarray}\label{eq:a1}
(\gamma_{\mu\nu} p_\mu p_\nu +m^2 )\psi_1 &=& 0\quad,\\ \label{eq:a2}
(\gamma_{\mu\nu} p_\mu p_\nu
-m^2 ) \psi_2 &=&0 \quad,
\end{eqnarray}
which possess solutions with  a correct physical dispersion.
The second equation  coincides with the Ahluwalia {\it et al.} equation  for
$v$ spinors (Eq. (13) ref.~[1b]) or with Eq. (12) of
ref.~[16c].\footnote{For the discussion of
differences between ref.~\cite{Ahl} and our
model  see ref.~\cite{Dvoegl2} and  what follows.}

If accept the concept of the Weinberg field as a set of
degenerate states, one has to allow for possible
transitions $\psi_1 \leftrightarrow \psi_2$ (or  $F_{\mu\nu} \leftrightarrow
\tilde F_{\mu\nu}$). Therefore, one can propose the Lagrangian
with the following dynamical part:
\begin{equation}\label{eq:lagran2}
{\cal L}^{(2)} = -
\partial_\mu \overline \psi_1 \gamma_{\mu\nu} \partial_\nu \psi_2
-\partial_\mu \overline \psi_2 \gamma_{\mu\nu} \partial_\nu \psi_1  \quad.
\end{equation}
But this form appears not to admit a mass term in a usual manner.
Prof. Sachs proposed to consider inertial mass as a varying parameter.
In the present framework we are going to go further: to consider
the possibility of the {\it complex} mass parameter. This may be
understood by means of the definition of the mass as the normalization
coefficient. The field operator may be defined in different forms
(the commutation relations as well) and different global phase factors
between fields $\psi_1$ and $\psi_2$ can lead to the normalization
changes. If consider $m^2$ as a pure imaginary quantity
($m = (1\pm i) \tilde m$), {\it i.~e.} in quantum theory
as an anti-Hermitian operator, one can write the mass terms
for the Lagrangian (\ref{eq:lagran2}). However, the problem of origin of
the mass term will be analyzed more carefully elsewhere.

At this moment  I have to treat the question of solutions
in the momentum space.
The explicit form of Hammer-Tucker bispinors, ref.~\cite{Tucker} (see
also refs.~\cite{Dvoegl,Novozh}), is
\begin{eqnarray}\label{b1}
u_1^\sigma (\vec p)= v_1^\sigma
(\vec p) =\frac{1}{\sqrt{2}}\pmatrix{\left [1+ {(\vec J\vec p)\over  m}
+{(\vec J \vec p)^2 \over  m(E+m)}\right ]\xi_\sigma \cr \left [ 1  -
{(\vec J\vec p)\over  m} +{(\vec J \vec p)^2 \over  m(E+m)}\right ]
\xi_\sigma}\quad,
\end{eqnarray}
for the equation (\ref{eq:a1}); and
\begin{eqnarray}\label{b2}
u_2^\sigma (\vec p)= v_2^\sigma (\vec p)
=\frac{1}{\sqrt{2}}\pmatrix{\left [1+
{(\vec J\vec p)\over  m} +{(\vec J \vec p)^2 \over  m(E+m)}\right
]\xi_\sigma \cr
\left [ - 1  + {(\vec J\vec p)\over  m} - {(\vec J \vec p)^2
\over  m(E+ m)}\right ] \xi_\sigma}\quad,
\end{eqnarray}
for the equation (\ref{eq:a2}).
The bispinor normalization in the cited papers is chosen as
$\overline u_1^\sigma (\vec p) u_1^\sigma (\vec p) =
\overline v_1^\sigma (\vec p) v_1^\sigma (\vec p) =
- \overline u_2^\sigma (\vec p) u_2^\sigma (\vec p) =
- \overline v_2^\sigma (\vec p) v_2^\sigma (\vec p) = 1$,
what  corresponds to   a normalization of   the ``Pauli
spinors" to $\xi^*_\sigma \xi_{\sigma^\prime} =\delta_{\sigma\sigma^\prime}$.

Using the plane-wave expansion,
{\it e.g.},~ref.~[1a,for\-mu\-la (8)] or Eqs. (\ref{pl1},\ref{pl2})
of the present paper, it is easy to convince ourselves
that both $u_i^\sigma$ and $v_i^\sigma$ satisfy
the equations
\begin{equation}\label{eq:sp1}
\left [ - \gamma_{44} E^2 +2iE\gamma_{4i} \vec p_i +\gamma_{ij}
\vec p_i \vec p_j  +
m^2\right ] u_1^\sigma (\vec p) = 0
\quad (\mbox{or}\,\,v_1^\sigma (\vec p)) \quad,
\end{equation}
and
\begin{equation}\label{eq:sp2}
\left [ - \gamma_{44} E^2 +2iE\gamma_{4i} \vec p_i +\gamma_{ij}
\vec p_i \vec p_j  -
m^2\right ] u_2^\sigma (\vec p) = 0
\quad (\mbox{or} \,\,v_2^\sigma (\vec p))\quad,
\end{equation}
respectively.

Bispinors of Ahluwalia {\it et al.}, ref.~\cite{Ahl2}, can be written
in a  more compact  form:
\begin{eqnarray}
u^\sigma_{AJG} (\vec p) =
\pmatrix{\left [ m +\frac{(\vec J \vec p)^2}{E+m}
\right ] \xi_\sigma\cr
(\vec J \vec p) \xi_\sigma}\quad, \quad
v^\sigma_{AJG} (\vec p) =\pmatrix{0 & 1\cr
1 & 0} u^\sigma_{AJG} (\vec p)\quad.
\end{eqnarray}
They coincide with the Hammer-Tucker-Novozhilov bispinors
within a normalization and an unitary transformation by
${\cal U}$ matrix:
\begin{eqnarray}
u^\sigma_{AJG} (\vec p) = m\,\,\, \cdot  {\cal U} u^\sigma_1 (\vec p) =
\frac{m}{\sqrt{2}}\pmatrix{1 & 1 \cr
1 & -1} u^\sigma_1 (\vec p)\quad,
\end{eqnarray}
\begin{eqnarray}
v^\sigma_{AJG} (\vec p) = m\,\,\, \cdot {\cal U} v^\sigma_2 (\vec p) =
\frac{m}{\sqrt{2}}\pmatrix{1 & 1 \cr
1 & -1} v^\sigma_2 (\vec p) \quad.
\end{eqnarray}

In the case of a choice $u_1^\sigma$ and $v_2^\sigma$ as
physical bispinors\footnote{I don't agree with the claim of the authors
of ref.~[1a,footnote 4] which states
 $v_1^\sigma (\vec p)$ are not solutions of the equation (\ref{eq:a1}).
 The origin of the possibility that the $u_i$-
and $v_i$- bispinors  in  Eqs. (\ref{eq:sp1},\ref{eq:sp2})
coincide  each other
(see  Eqs. (\ref{b1},\ref{b2})) is the following: the Weinberg equations
are of the second order in  time derivatives.
The detailed analysis will be presented in future publications.

In the meantime, I agree that it is
more convenient to work with bispinors normalized to
the mass, {\it e.g.}, $\pm m^{2j}$. In the following I keep
the normalization of bispinors as in ref.~\cite{Ahl}.}
we come to the
Bargmann-Wightman-Wigner-type (BWW) quantum field model proposed
by Ah\-lu\-wa\-lia {\it et al.} Of course, following  to the
same logic one can choose $u_2^\sigma$ and $v_1^\sigma$
bispinors and come to  the reformulation of the BWW theory.
Though in this case parities of a boson and
its antiboson are opposite, we have $-1$ for $u$ bispinor and
$+1$ for $v$ bispinor, {\it i.e.} different in the sign from
the model of Ahluwalia {\it et al.}, ref.~\cite{Ahl}.

Now let me repeat the quantization procedure of ref.~\cite{Hayashi},
however, it will be applied to the Weinberg field. Let me trace
the contributions of ${\cal L}^{(1)}$ and ${\cal L}^{(2)}$ to  dynamical
invariants separately .

{}From the definitions~\cite{Lurie}:
\begin{eqnarray}
{\cal T}_{\mu\nu} &=& -\sum_i \left \{
\frac{\partial {\cal L}}{\partial (\partial_\mu \phi_i )}
\partial_\nu \phi_i
+ \partial_\nu \overline \phi_i\frac{\partial {\cal L}}{\partial
(\partial_\mu \overline \phi_i )} \right \}
+{\cal L}\delta_{\mu\nu}\quad,\\
P_\mu &=& \int {\cal P}_{\mu} (x) d^3 x = -i\int {\cal T}_{4\mu} d^3 x
\end{eqnarray}
one can find  the energy-momentum tensor
\begin{eqnarray}
\lefteqn{{\cal T}^{(1)}_{\mu\nu} = \partial_\alpha \overline \psi_1
\gamma_{\alpha\mu} \partial_\nu \psi_1 + \partial_\nu \overline \psi_1
\gamma_{\mu\alpha} \partial_\alpha \psi_1 +\nonumber}\\
&+&\partial_\alpha \overline \psi_2
\gamma_{\alpha\mu} \partial_\nu \psi_2 +\partial_\nu \overline \psi_2
\gamma_{\mu\alpha} \partial_\alpha \psi_2+{\cal L}^{(1)}\delta_{\mu\nu}
\quad,
\end{eqnarray}
and
\begin{eqnarray}
\lefteqn{{\cal T}_{\mu\nu}^{(2)} = \partial_\alpha \overline\psi_1
\gamma_{\alpha\mu} \partial_\nu  \psi_2 + \partial_\nu \overline \psi_1
\gamma_{\mu\alpha} \partial_\alpha \psi_2 + \nonumber}\\
&+& \partial_\alpha \overline \psi_2 \gamma_{\alpha\mu} \partial_\nu \psi_1
+\partial_\nu  \overline \psi_2 \gamma_{\mu\alpha}\partial_\alpha \psi_1 +
{\cal L}^{(2)} \delta_{\mu\nu}\quad.
\end{eqnarray}

As a result the first part of the Hamiltonian
 is\footnote{The Hamiltonian  can also be obtained from
the second order Lagrangian presented in~[1b,Eq. (18)] by means of the
procedure
developed by  M. V. Ostrogradsky~\cite{Ostrog} long ago (see also
Weinberg's remark
on the page B1325 of the first paper~\cite{Weinberg}). However,
it would be difficult me to agree with the definition of
momentum conjugate operators in the paper~[1a].
The Ostrogradsky's procedure seems not to have been applied
there to obtain momentum conjugate operators.}
\begin{eqnarray}\label{H}
\lefteqn{{\cal H}^{(1)} = \int \left [ -\partial_4 \overline \psi_2
\gamma_{4 4}\partial_4 \psi_2 +
\partial_i \overline \psi_2  \gamma_{ij} \partial_j \psi_2
-\right.\nonumber}\\
&-& \left.  \partial_4 \overline \psi_1 \gamma_{4 4} \partial_4 \psi_1 +
\partial_i \overline \psi_1  \gamma_{ij} \partial_j \psi_1
+m^2 \overline \psi_1 \psi_1
-m^2 \overline \psi_2 \psi_2 \right ] d^3 x \quad,
\end{eqnarray}
and the second part is
\begin{eqnarray}\label{H2}
\lefteqn{{\cal H}^{(2)} = \int \left [ -\partial_4 \overline
\psi_2 \gamma_{4 4} \partial_4 \psi_1 +
\partial_i \overline \psi_2  \gamma_{ij} \partial_j \psi_1
-\right.\nonumber}\\
&-& \left.  \partial_4 \overline \psi_1 \gamma_{4 4} \partial_4 \psi_2 +
\partial_i \overline \psi_1  \gamma_{ij} \partial_j \psi_2\right ] d^3 x
\quad.
\end{eqnarray}

Using the plane-wave expansion
\begin{eqnarray}\label{pl1}
\psi_1 (x) &=&\sum_\sigma \int \frac{d^3 p}{(2\pi)^3} \frac{1}{m \sqrt{2E_p}}
\left [ u_1^\sigma (\vec p) a_\sigma (\vec p) e^{ipx} +v_1^\sigma (\vec p)
b^\dagger_\sigma (\vec p) e^{-ipx} \right ]\quad,\\
\label{pl2}
\psi_2 (x) &=&\sum_\sigma \int \frac{d^3 p}{(2\pi)^3} \frac{1}{m\sqrt{2E_p}}
\left [ u_2^\sigma (\vec p) c_\sigma (\vec p) e^{ipx} +v_2^\sigma (\vec p)
d^\dagger_\sigma (\vec p) e^{-ipx} \right ] \quad,
\end{eqnarray}
$E_p=\sqrt{\vec p^{\,2} +m^2}$,
one can come to the quantized Hamiltonian
\begin{equation}
{\cal H}^{(1)} =  \sum_\sigma \int \frac{d^3 p}{(2\pi)^3} E_p \, \left [
a_\sigma^\dagger (\vec p) a_\sigma (\vec p) +b_\sigma (\vec p) b_\sigma^\dagger
(\vec p) +c_\sigma^\dagger (\vec p) c_\sigma(\vec p) +d_\sigma (\vec p)
d^\dagger_\sigma (\vec p)\right ]\,,
\end{equation}
\begin{equation}
{\cal H}^{(2)} =  \sum_\sigma \int \frac{d^3 p}{(2\pi)^3} E_p \, \left [
a_\sigma^\dagger (\vec p) c_\sigma (\vec p) +b_\sigma (\vec p) d_\sigma^\dagger
(\vec p) +c_\sigma^\dagger (\vec p) a_\sigma(\vec p) +d_\sigma (\vec p)
b^\dagger_\sigma (\vec p)\right ]\, ,
\end{equation}
following the procedure of, {\it e.g.}, refs.~\cite{Bogol,Itzyk}.

Setting up  boson commutation relations  as follows:
\begin{eqnarray}\label{c1}
\left [a_\sigma (\vec p)+c_\sigma (\vec p), a^\dagger_{\sigma^\prime} (\vec
k)+c^\dagger_{\sigma^\prime} (\vec k) \right ]_{-} &=& (2\pi)^3
\delta_{\sigma\sigma^\prime}\delta (\vec p - \vec  k)\quad,\\ \label{c2}
\left [b_\sigma (\vec p)+d_\sigma (\vec p), b^\dagger_{\sigma^\prime}
(\vec  k) +d^\dagger_{\sigma^\prime} (\vec k) \right ]_{-} &=& (2\pi)^3
\delta_{\sigma\sigma^\prime}\delta (\vec p - \vec  k)\quad,
\end{eqnarray}
it is easy to see that the Hamiltonian is  positive-definite and
the translational invariance still keeps  in the framework of this
description ({\it cf.} with ref.~\cite{Ahl}). Please pay attention here:
there is no indefinite metric involved.

Analogously, from the definitions
\begin{eqnarray}
{\cal J}_{\mu} &=& -i \sum_i \left \{ \frac{\partial {\cal L}}{\partial
(\partial_\mu \phi_i)}
\phi_i -\overline \phi_i \frac{\partial {\cal L}}{\partial
(\partial_\mu \overline\phi_i)} \right \}\quad, \\
Q &=& -i \int {\cal J}_{4} (x)  d^3 x \quad,
\end{eqnarray}
and
\begin{eqnarray}\label{PL}
\lefteqn{{\cal M}_{\mu\nu,\lambda} = x_\mu {\cal T}_{\lambda\nu}
- x_\nu {\cal T}_{\lambda\mu} - \nonumber}\\
&-&i \sum_{i}
\left \{ \frac{\partial {\cal L}}
{\partial (\partial_\lambda \phi_i)} N_{\mu\nu}^{\phi_i} \phi_i +
\overline \phi_i  N_{\mu\nu}^{\overline\phi_i} \frac{\partial {\cal L}}
{\partial (\partial_\lambda \overline \phi_i)}\right \} \quad,\\
M_{\mu\nu} &=& -i \int  {\cal M}_{\mu\nu, 4} (x) d^3 x  \quad,
\end{eqnarray}
I found the current operator
\begin{eqnarray}\label{qq1}
\lefteqn{{\cal J}^{(1)}_\mu = i \left [\partial_\alpha \overline \psi_1
\gamma_{\alpha\mu}
\psi_1 - \overline \psi_1 \gamma_{\mu\alpha} \partial_\alpha  \psi_1
+\right.\nonumber}\\
&+&\left.  \partial_\alpha\overline \psi_2 \gamma_{\alpha\mu}  \psi_2
- \overline \psi_2
\gamma_{\mu\alpha} \partial_\alpha \psi_2\right ]\quad,
\end{eqnarray}
and
\begin{eqnarray}\label{qq2}
\lefteqn{{\cal J}^{(2)}_\mu = i \left [\partial_\alpha \overline \psi_2
\gamma_{\alpha\mu}
\psi_1 - \overline \psi_2 \gamma_{\mu\alpha} \partial_\alpha  \psi_1
+\right.\nonumber}\\
&+&\left.  \partial_\alpha\overline \psi_1 \gamma_{\alpha\mu}  \psi_2 -
\overline \psi_1 \gamma_{\mu\alpha} \partial_\alpha \psi_2\right ]\quad.
\end{eqnarray}

Using (\ref{PL})  the spin momentum tensor reads
\begin{eqnarray}\label{ss1}
\lefteqn{S_{\mu\nu,\lambda}^{(1)} = i \left [ \partial_\alpha \overline \psi_1
\gamma_{\alpha\lambda} N_{\mu\nu}^{\psi_1} \psi_1 +  \overline \psi_1
N_{\mu\nu}^{\overline \psi_1}\gamma_{\lambda\alpha}
\partial_{\alpha} \psi_1 + \right.\nonumber}\\
&+& \left. \partial_\alpha \overline \psi_2
\gamma_{\alpha\lambda} N_{\mu\nu}^{\psi_2} \psi_2
+\overline \psi_2 N_{\mu\nu}^{\overline \psi_2}
\gamma_{\lambda\alpha} \partial_\alpha \psi_2 \right ]\quad,
\end{eqnarray}
\begin{eqnarray}\label{ss2}
\lefteqn{S_{\mu\nu,\lambda}^{(2)} = i \left [ \partial_\alpha \overline \psi_2
\gamma_{\alpha\lambda} N_{\mu\nu}^{\psi_1} \psi_1 +  \overline \psi_2
N_{\mu\nu}^{\overline \psi_2}\gamma_{\lambda\alpha}
\partial_{\alpha} \psi_1 + \right.\nonumber}\\
&+& \left. \partial_\alpha \overline \psi_1
\gamma_{\alpha\lambda} N_{\mu\nu}^{\psi_2} \psi_2
+ \overline \psi_1 N_{\mu\nu}^{\overline \psi_1} \gamma_{\lambda\alpha}
\partial_\alpha \psi_2 \right ]\quad.
\end{eqnarray}

If  the Lorentz group generators  (a $j=1$ case) are defined from
\begin{eqnarray}\label{lor}
&&\overline  \Lambda \gamma_{\mu\nu} \Lambda a_{\mu\alpha}
a_{\nu\beta} = \gamma_{\alpha\beta}\quad,\\
&& \overline \Lambda \Lambda =1\quad,\\ \label{lor3}
&&\overline \Lambda  = \gamma_{44} \Lambda^\dagger \gamma_{44}\quad.
\end{eqnarray}
then in order to keep the  Lorentz covariance of
the Weinberg
equations and of the Lagrangian (\ref{eq:Lagran1})
one can use the following generators:
\begin{equation}
N_{\mu\nu}^{\psi_1,  \psi_2 (j=1)} =
- N_{\mu\nu}^{\overline \psi_1 , \overline \psi_2 (j=1)} =
{1\over 6} \gamma_{5,\mu\nu}\quad,
\end{equation}
see also ref.~[16b, Eqs. (37,51,52)].
The matrix  $\gamma_{5,\mu\nu}= i \left [\gamma_{\mu\lambda},
\gamma_{\nu\lambda}\right ]_{-}$ is  defined to be Hermitian.

The quantized charge operator and the quantized  spin
operator follow immediately from (\ref{qq1},\ref{qq2}) and (\ref{ss1}):
\begin{equation}
Q^{(1)}=\sum_\sigma \int \frac{d^3 p}{(2\pi)^3} \, \left [
a_\sigma^\dagger (\vec p) a_\sigma (\vec p) -
b_\sigma (\vec p) b_\sigma^\dagger (\vec p) +
c_\sigma^\dagger (\vec p) c_\sigma (\vec p) -
d_\sigma (\vec p) d_\sigma^\dagger (\vec p)\right ]\,,
\end{equation}
\begin{equation}
Q^{(2)} =  \sum_\sigma \int \frac{d^3 p}{(2\pi)^3}\, \left [
a_\sigma^\dagger (\vec p) c_\sigma (\vec p) -
b_\sigma (\vec p) d_\sigma^\dagger (\vec p) +
c_\sigma^\dagger (\vec p) a_\sigma (\vec p) -
d_\sigma (\vec p) b_\sigma^\dagger (\vec p)\right ]\, ,
\end{equation}
\begin{eqnarray}\label{spin}
\lefteqn{(W^{(1)}\cdot n) =   \sum_{\sigma\sigma^\prime}\int \frac{d^3
p}{(2\pi)^3}{1\over m^2 E_p} \overline u_1^\sigma (\vec p) (E_p \gamma_{44} -
i\gamma_{4i}p_i )\,\, I \otimes (\vec J \vec n)\,  u^{\sigma^\prime}_1
(\vec p) \times \nonumber}\\ & &\qquad \times \left [a_\sigma^\dagger (\vec p)
a_{\sigma^\prime} (\vec p) + c_\sigma^\dagger (\vec p) c_{\sigma^\prime} (\vec
p) - b_\sigma (\vec p) b_{\sigma^\prime}^\dagger (\vec p) -d_\sigma  (\vec p)
d_{\sigma^\prime}^\dagger  (\vec p) \right ]
\end{eqnarray}
(provided that  the frame is chosen in such a way
that $\vec n \,\,\, \vert\vert \,\,\,\vec p$ \,\,\, is along the third axis).
It is easy to verify  the eigenvalues
of the charge operator\footnote{In order to construct  neutral particle
operators one can
use  an analogy with a $j=1/2$ case (to compare electron and neutrino
field operators). At
the present moment I would like again draw your attention to the fact that
$u_i^{(j)}$ and $v_i^{(j)}$ coincide in the model studied here.} are $\pm
1$, and  of the Pauli-Lubanski  spin operator are
\begin{equation}\label{heli}
\xi^*_\sigma (\vec J \vec n) \xi_{\sigma^\prime} = +1,\, 0\, -1
\end{equation}
in a massive case and $\pm 1$
in a massless case (see the discussion on the massless limit of the Weinberg
bispinors in ref.~\cite{Ahl2}).

As for the spin operator which follows from the Lagrangian (\ref{eq:lagran2})
the situation is more difficult.  If  accept
(\ref{lor})-(\ref{lor3})
we are not able to  obtain the helicity operator  (\ref{heli})
in the final expression. However, a cure is possible. We should take
into account  that the transformations
\begin{eqnarray}
&&\overline \Lambda \gamma_{\mu\nu} \Lambda a_{\mu\alpha} a_{\nu\beta} = \mp
\gamma_{\alpha\beta}\gamma_5\\
&&\overline \Lambda \Lambda = \pm \gamma_5\quad,
\end{eqnarray}
remain  Weinberg's set of equations to be invariant.
Equations are only interchanged each other.  This is a cause of
the possibility of  combining the Lorentz and the dual (chiral)
transformation for the Weinberg degenerate doublet.
Thus, in order to  keep the  Lorentz and parity covariance of
 the Lagrangian of the form (\ref{eq:lagran2}) one can impose
\begin{equation}
N_{\mu\nu}^{\psi_1,  \psi_2 (j=1)} =
- N_{\mu\nu}^{\overline \psi_1 , \overline \psi_2 (j=1)} =
{1\over 6} \gamma_{5,\mu\nu}\gamma_5\quad.
\end{equation}

After simple transformations I obtain
\begin{eqnarray}\label{spin1}
(W^{(2)}\cdot n) &=& \sum_{\sigma\sigma^\prime}\int \frac{d^3 p}{(2\pi)^3}
\xi^*_\sigma (\vec J \vec n) \xi_{\sigma^\prime}  \times\nonumber\\
&\times& \left [c_\sigma^\dagger
(\vec p)
a_{\sigma^\prime} (\vec p) + a_\sigma^\dagger (\vec p) c_{\sigma^\prime}
(\vec p) - b_\sigma (\vec p) d_{\sigma^\prime}^\dagger (\vec p) -d_\sigma
(\vec p) b_{\sigma^\prime}^\dagger  (\vec p) \right ]\quad.
\end{eqnarray}
I  leave  investigations of
other possibilities for  further publications.

Why  ``a queer reduction of degrees of freedom"
did happen  in the previous papers~\cite{Avdeev,Avd2,Hayashi,Kalb}?
The origin of this surprising fact  follows from Hayashi's
(1973) paper, ref.~\cite[p.498]{Hayashi}:   The requirement  of
{\it ``that the physical realizable state
satisfies  a quantal version of the generalized Lorentz condition"},
formulas (18) of ref.~\cite{Hayashi},\footnote{Read:  ``a quantal
version" of the  Maxwell equations imposed
on state vectors in the Fock space.   See
the papers  of Ahluwalia {\it et al.},
{\it e.g.}, ref.~\cite[Table 2]{Ahl3}, for  the discussion
on the  acausal physical dispersion of the
equations (4.19) and (4.20) of ref.~[11b],
``which {\it are just Maxwell's free-space equations for left-
and right- circularly polarized radiation."} See also
the footnote \# 1 in ref.~\cite{Dvoegl2} and ref.~[3g].
Let me  mention that this fact is probably connected  with the
indefinite metric problem.}  permits one
to eliminate  upper (or down) part of the Weinberg bispinor  and
to remove transversal components of the remained part
by means of the ``gauge" transformation  (\ref{eq:gauge}),
what {\it ``ensures the massless skew-symmetric
field is longitudinal".}
The reader  can convince himself of this obvious fact  by looking
at the explicit form of
the Pauli-Lubanski operator, Eq. (\ref{spin}).
Keeping all terms in field operators and
in the Lagrangian (\ref{eq:Lagran}), {\it cf.} with~\cite{Avdeev,Avd2},
and {\it not} applying
the generalized Lorentz condition, {\it cf.} with~\cite{Hayashi,Kalb},
we are able  to obtain transversal components, {\it i.e.},
a $j=1$ particle in the massless  limit of the Weinberg theory.
The presented version does {\it not} contradict with
the Weinberg theorem nor with the classical limit, Eqs. (21,22) of
ref.~\cite{Dvoegl2}. Thanks to the mapping~[3e,4]
the conclusion is valid for both the Weinberg  $2(2j+1)$
component ``bispinor" and the antisymmetric
(skew-symmetric) tensor  field.

Finally, for the sake of completeness let me re-write the Lagrangians
presented above  in the form:
\begin{equation}
{\cal L}_1  = - \partial_\mu \overline \Psi \Gamma_{\mu\nu} \partial_\nu \Psi
- m^2 \overline \Psi \Psi \quad,
\end{equation}
where
\begin{eqnarray}
\Psi = \pmatrix{\psi_1 \cr \psi_2}\quad,
\quad \overline \Psi = \pmatrix{\psi_1^\dagger & \psi_2^\dagger \cr}
\cdot \pmatrix{\gamma_{44}&0 \cr   0& -\gamma_{44}}
\end{eqnarray}
are the wave function of the degenerate doublet;
\begin{eqnarray}
\Gamma_{\mu\nu} = \pmatrix{\gamma_{\mu\nu} &0 \cr 0& -\gamma_{\mu\nu}\cr}
\quad.
\end{eqnarray}
The Lagrangian ${\cal L}_2$ could be written in a similar form if imply
\begin{eqnarray}
\overline \Psi = \pmatrix{
\psi_1^\dagger &  \psi_2^\dagger  \cr}
\cdot
\pmatrix{
0 &  -\gamma_{44}\cr
\gamma_{44} & 0 \cr
} \quad.
\end{eqnarray}

I take the liberty to name
the field operator $\overline \Psi$  as the  Weinberg-conjugated
{\it dibispinor}.

My conclusions are:  There exist the versions of   both the Weinberg
$2(2j+1)$ component theory and the antisymmetric tensor field formalism
that answer for particles with  transversal components. Thus, these
versions do not contradict the Weinberg theorem and, in the case of
$j=1$ particle, the doublet field $(\psi_1,\quad \psi_2)$, or
$(F_{\mu\nu},\quad \tilde F_{\mu\nu})$, could be used in describing
a photon. The origin of the contradictions met in the previous papers
is the inadequate use of the generalized Lorentz condition
which may be incompatible (in such a form) with the specific properties of
antisymmetric tensor fields. The connection of the present model
with the Bargmann-Wightman-Wigner-type quantum field theories
deserves further elaboration.

As a matter of fact the present model develops Weinberg's and Ahluwalia's
ideas of the Dirac-like description of bosons on an equal footing with
fermions, {\it i.e.}, on the basis of the $(j,0)\oplus (0,j)$
representation of the Lorentz group.

\smallskip

{\bf Acknowledgments.} The papers, letters and sincere
discussions with  Prof. ~D.~V.~Ahluwalia
were  very  helpful  in writing of this paper.
Many thanks to him. Discussions with Profs. I.~G.~Kaplan, M.~Moreno,
Yu.~F.~Smirnov and M.~Torres were very  useful.

I am grateful to Zacatecas University for a professorship.

\end{document}